\documentclass{article}
\usepackage{amsmath}
\usepackage{amsthm}
\usepackage{amssymb}
\usepackage{hyperref}
\usepackage[affil-it]{authblk}

\def\R{\ensuremath\mathbb R}

\def\1{\ensuremath\mathbf 1}

\begin{document}

\author{Salil Bhate}
\affil{ Department of Bioengineering,\\ Stanford Unviversity}
\title{Theoretical foundations for the Human Cell Atlas}
\maketitle

\section*{Abstract}
In Schiebinger et al. (2017), the authors use optimal transport of measures on empirical distributions arising from biological experiments to relate the single cell RNA sequencing profiles for induced pluripotent stem cells differentiating. But such algorithms could be arbitrarily applied to any datasets from any collection of experiments. We consider here a natural question that arises: in a manner consistent with conventionally accepted assumptions about biology, in which cases can the results of two experiments be mapped to each other in this manner?  The answer to this question is of fundamental \emph{practical} importance in developing algorithms that use this method for analysing and integrating complex datasets collected as part of  the Human Cell Atlas.

Here, we develop a formulation of biology in terms of sheaves of $C^*(X)$-modules for a smooth manifold $X$ equipped with certain structures,  that enables this question to be formally answered, leading to formal statements about \emph{experimental inference} and \emph{phenotypic identifiability}.  These structures capture a perspective on biology that is consistent with a standard, widely accepted biological perspective and is mathematically intuitive.

The statement reagrding identifiability of phenotypes is, informally:

\begin{center}
 \emph{ Given an organism in a given condition, a phenotype is identifiable if there exists an experiment that identifies it}.
 \end{center}
 
 The statement regarding experimental inference can is, informally:
 \begin{center}
\emph{ Given an organism, that is comprised of a collection of tissues, optimal transport of measures can be consistently used to relate the results of experiments on each tissue if and only if there exists an experiment design on the whole organism that restricts to each of the experiment designs on the tissue.}
\end{center}

Our methods provide a framework in which to design complex experiments and the algorithms to analyse them in a way that their conclusions can be believed.

\section*{Introduction}
Biological organisms are very complex systems. In spite of this, for hundreds of years, biologists have built up a theory of how these systems work. Their success has required them to conceptualize and compartmentalize these systems in a way that enables them to make progress, in spite of their overbearing complexity. However, a biologist would struggle to articulate in precise mathematical terms how they do this.

For years, statisticians have developed tools that have found numerous real-world applications. Geometers in contrast, can precisely articulate the strategies they use to make progress in the face of overbearing complexity, but many would struggle to find applications. It turns out, these strategies can be directly used to describe biological systems, in a way that is consistent with how biologists do it themselves.

In section 1, we show that the tools developed by geometers are the precise ones needed to articulate the complexity of biological systems. This section is for mathematicians to conceptualize how biologists think about biology.  This leads to very natural constructions, that have probably been developed in their most abstract form elsewhere in the mathematical literature. The goal is to develop a mathematical description for  biological systems that is consistent with the way biologists have been conceptualizing these systems all along.  

In section 2, we develop a the mathematical formalism that describes the bridge between this theory and the process of biological experimentation. Biologists won't understand these first two sections, but the names for constructions will make it clear how the mathematical contexts map to the ones they're familiar with.

In section 3, we make the statements regarding inference in biological systems and phenotypic identifiability.

Our formulation provides the framework to ask theoretical questions in a biologically relevant way. In addition, it provides a framework in which novel statistical, algorithmic and physical tools can be developed in order to make predictions about emergent properties of biological systems that we are not yet able to measure. 

\section{How biologists think about biology}
The constructions in this section may seem very unnatural to mathematicians: why are we thinking about these spaces at all? Aren't all the sheaves soft? Why do we need to think about sheaves instead of just their sections, or presheaves? The answer is because: the language of sheaves precisely captures notions that are very intuitive to most biologists, but that they cannot formally define. The nomenclature used should provide a direct mapping between a biological concept and the mathematical construction.

\subsection{Molecules}
The collection of molecules $M$ is a discrete set of all molecules that we're considering

\subsection{Definition: organism}
An organism $X \subset \R^3\times [0,1]$ is a smooth manifold together with the  sheaf $\mathcal F = C^{\infty}(X,[0,1]^{|M|})$, considered as a $C^*(X)$ module. Each section is smooth probability distribution function at each point for a random variable taking values on $\{0,1\}^{|M|}$. This corresponds to the probability of finding any given combination of molecules at position $x$ and time $t$ for  $(x,t) \in X \subset \R^3$. 

We call $\mathcal F$ an organism's condition sheaf, and a section of it we call a condition.

\subsection{Definition: tissue}
For  a smooth open submanifold $Y \subset X$, a tissue is the organism obtained by restriction $(X|_Y, \mathcal F|_Y)$

\subsection{Definition: phenotype}
For a smooth Euclidean  manifold $K$, any map $ f \in C^*(X \times \{0,1\}^{|M|}) \to K$ induces a morphism of sheaves, defined on sections by
\[ T_f : p \mapsto \int f(x,m) dp(m)\]

Such a map $f$ is a phenotype, and $T_f$ is maps a condition to a distribution over phenotypes. This corresponds to defining at each point in $X$ a probability distribution over $K$

\subsection{Interpretation}
\begin{enumerate}

\item 
An organisms condition defines, for each point in it (in space and time), the likely combination of all the arrangements of all the molecules

\item
An organism's condition sheaf corresponds to all the possible conditions an organism could possibly have.

\item
A phenotype corresponds to \emph{some function} that depends deterministically on the arrangements of all the molecules across space and time.

\end{enumerate}

One might ask why we're thinking about everything as sheaves and not just as functions. The reason is to capture the notion of \emph{emergent phenotypes}. This term is typically used to refer to phenotypes that are more than just the sum of the phenotypes on each each tissue; it's clear why sheaves are the appropriate tools to ask such questions.

\subsection{Definition: phenotype class}
A sub sheaf of $C^*(X\times \{0,1\}^{|M|}, K)$ is called a $K$-valued phenotype class.

\subsection{Interpretation of phenotype classes}
While we want our approaches to be consistent if we were to conceptualize biological systems in their greatest possible granularity, in terms of all the molecules, we don't want to think of phenotypes as depending on all the possible combinations of all the possible molecules in space and time, because we can't do any experiments if that's the case. We want to group them together in meaningful ways and decide which experiments to do; for example, some things, like genetic conditions, are phenotypes that are observed in a time-independent way. Other things, like somebody's age, don't depend on space. Other things, like heart-rate, depend on time and space but only in the blood.

For any measurable $Y \subset X\times \{0,1\}^{|M|}$, we can think of the $K$-phenotype class corresponding to functions that vanish on $Y$. 
\begin{itemize}
\item
We can think of a cell $C$ as a smooth cell complex contained in $X$ that is a submanifold. So for any cell, we can define the $K$-phenotype class of functions that are non-zero only on $C$.
\item 
Similarly, we can define the $K$-phenotype class of phenotypes that are continuous on the union $X = \bigcup C_i$ where $C_i$ are a collection of cells. This is the $K$-phenotype class of 'single-cell' phenotypes.
\item
We can similarly define a $K$-phenotype class of functions for every tissue; these are the tissue phenotypes.
\item
For any collection of molecules $S \subset M$, we can think of the $K$-phenotype class of  phenotypes that are nonzero only on $X \times \{S\}$. For example, if $S$ is the collection of RNA-molecules, get the $K$-phenotype class of transcriptional phenotypes
\item We can think of the phenotype class given by functions with no dependence on $X$, these are global phenotypes. Similarly, we can think of time invariant phenotypes.
\end{itemize}

Different research labs are interested in different phenotype classes:
\begin{itemize}
\item
Immunologists typically consider the phenotype class of functions that depend only on the surface of a cell and ~200 protein molecules called CD markers, cluster of differentiation markers.
\item
The term `systems biology' corresponds to considering the phenotype class of phenotypes that depend on multiple tissues.

\item
Most labs consider phenotype classes that are smooth in time but that doesn't mean that all phenotypes that are biologically observed are. Recently, phenomena like \emph{transcriptional bursting} have been described.

\end{itemize}

\section{Biological measurements}
When biologists say they're interested in some area of biology, they mean that they're interested in an organism $X$, a sub-sheaf of its condition-sheaf as well as some phenotype class. A biologist is interested in a phenotype class, but that doesn't mean that they are able to easily observe it: they observe it indirectly, by means of an experiment. A measurement could be something like
\begin{itemize}
\item
Measuring a mouse's heart rate, a single number
\item 
Subjecting it to a treatment; for example seeing if a monkey chooses a banana or an apple in a maze, a measurement valued in \{apple, banana\}
\item 
Measuring the transcription level of all of its genes, a list of around 25,000 numbers corresponding to each gene sequence.
\end{itemize}

But it typically consists of many of these. In general, these experiments are not necessarily deterministic, which is why we make the following definition.

\subsection{Definition: observation transformation}
For  smooth manifolds $K_1, K_2$, an observation transformation is a morphism $\phi$ of sheaves of $C^*(X)$ modules $\mathcal C^*(X,K_1) \to \mathcal C^*(X,K_2)$ 

\subsection{Experiment}
We say that an observation transformation is an experiment if $K_1 = \{0,1\}^{|M|}$ in the definition above, and  $\phi_*p$ is integrable for every condition an organism is in. The dimensionality of $K_2$ is the dimensionality of the experimental transformation.

\subsection{Experiment design}
Fix an organism $X$. An experiment design on $X$ is an open cover $X = \bigcup Y_i$, where $Y_i$ are tissues and  $E = (Y,\phi)$ where $Y$ is a tissue and $\phi$ are observation transformations.

\subsection{Interpretation}
 One might ask why we're thinking about everything as sheaves and not just as functions. The reason is to capture the notion of measurement consistency: if we measure the proteins in a whole brain, we should get the same answer as if we had cut it into pieces, however small, and measured the proteins within them. This is the obvious answer. Another reason is that in section 3, we'll talk about whether it's possible to reconstruct a measurement on a whole tissue given measurements on parts of it: this is a natural problem that arises. For example, single-cell RNAseq collects the RNA from individual cells in from a tissue but is a \emph{different} measurement technology to bulk RNAseq, which is used for measuring RNA; these correspond to different experiments in our framework! Something that the biological field is very actively interested problems of reconstructing the whole from \emph{different} experiments on each piece: these problems are precisely why we want to use the language of sheaves. Every sheaf doesn't always have the appropriate sections.
 
 Many biological experiments are complex, multi-stage experiments. For example, data is collected and computational algorithms process the data. An experiment, when composed with an observation transformation, is still an experiment.

\subsection{Definition: results of an experiment}
Given an organism $X$, a condition $p$ and an experiment design $\phi$, a biological sample is a collection of random samples valued in $K$ with probability distribution $\phi(p)$. These are actual numbers that a biologist obtains. The empirical  distribution that these provide can be written
\[ \sum_{\text{samples}} \delta_x \]
where $x$ denotes the value of the observation at a given sample.

\section{Inference in biological systems}

\subsection{Transport maps}
Suppose we are given a nonnegative kernel $C$ on euclidean spaces $C:K_1\times K_2 \to \R^+$; corresponding to some notion of distance. Then given any two distributions on $K$, we can always consider the optimal transport of measures between them with respect to $C$.

In particular, given any two empirical distributions (data from an experiment), we can consider the optimal transport mapping between them, which is something for which many fast algorithms exist and are being developed.

\subsection{Some examples of biological experiments}
Biologists are very familiar with thinking about transport maps between empirical distributions. Here are some examples.
\begin{enumerate}

\item In every clinical trial for a new drug, we have  patients. We collect the  information of each patient before they receive the treatment. We have a control group and a treatment group. We administer the treatment, and at the end of the trial, we collect the information. We transport the empirical distributions at the first time point to the second time point using the cost function. Let $x_0$ denote the data for a patient before the treatment, and $y$ the data for a patient after the treatment.
\[ C(\delta_x,\delta_y) = 0 \text{ if $x$ and $y$ came from the same patient and } \infty \text{ otherwise}\]

\item CAR-T cells are engineered cells that are injected into a patient with cancer. These have been recently approved in clinical trials. There is a lot of interest in understanding how they are behaving inside the body months after the treatment has begun. The way that biologists do this experiment is by measuring the T cells before injecting them, and taking a sample of the patients blood 6 months later. The blood includes T cells that aren't the injected  ones. So how do biologists map these? The distributions are mapped by means of a `CAR specific antibody':
\[ C(\delta_x, \delta_y) = 0 \text{ if $x$ and $y$ stain positively for the antibody and } \infty \text{ otherwise}\]

\item In a recent work in by Schiebinger et al (2017), experiments are collected corresponding to samples of single cell transcriptomes across a time course in differentiation. $x$ and $y$ are 25000 dimensional vectors. The cost function they used to map these distributions are:

\[ C(\delta_x, \delta_y) = K(x,y) \]
where $K$ is a diffusion kernel $K(x,y) =\exp^{\gamma||x-y||^2}$
\end{enumerate}

\subsection{A general purpose algorithm for making biological inferences}
Suppose we have the results of an experiment. We can always define a distance kernel, and consider the optimal transport of the empirical distribution defined by these, an obtain an answer for how one set of samples maps into the other. Of course, in some biological conditions, it's obvious whether the cost function is sensible, for example in the first two experiments. In other cases, it's not clear whether the cost function makes sense, a priori. It turns out that this cost function \emph{is} appropriate, for one reason: all of the cells between at one time point arise from the cells at the previous time point, and the gene expression is changing smoothly and this diffusion makes sense.

How do we precisely and mathematically relate when it makes sense to consider experiments in this way? In our formulation of biological experimentation:
\subsection{Statement about inference in biological experiments}

\emph{Given an organism, that is comprised of a collection of tissues $X = \bigcup U_i$, optimal transport of measures can be consistently used to relate the results of an experiment, with designs $E_i$ if and only if  there is an experimental design on $X$ which restrict to each of the experimental designs on the tissues.}

\subsection{Statement about the identifiability of phenotypes}
Given a phenotype $f$ on an organism $X$. We say that it is identifiable in condition $p$ if there exist tissues $U_i$ and  experiment designs $E_i = (U_i, \phi_i)$ on $U_i$ such that there exists an experiment design $E = (X, \phi)$ with the property that
\[  \int f d(\phi^* \mu) = \int f d(p)   \]
where $\mu$ is the lift of $\phi_* dp|_{U_i}$

We say that the collection of experiment designs $E$ identify $f$.

These definitions extend naturally to phenotype classes.

Note that this notion of identifiability gives rise to the opportunity to ask many questions of the form: given a measurement, what is the largest phenotype class that can be identified by it. Given a phenotype class, what measurements do we need in order to identify every phenotype within it. These questions can  be framed as theorems to prove; combinatorially, physically, statistically.

\section{Acknowledgements}
I would like to acknowledge Garry Nolan for his enthusiasm about developing experimental technologies that try and capture the multi-scale nature of biological systems, and his encouragement of diverse approaches to analyze data from these. It was Graham Barlow's constant repetition of  \emph{`biology is hierarchical'} that led to this project. Aleksander Doan alerted me to the fact that in an early version, it was unclear whether I was trying to explain mathematics to a biologist or biology to a mathematician, causing universal confusion. I would like to acknowledge Anna Seigal for mathematical discussions about this approach.

\section{References}
Schiebinger et al. (2017). Reconstruction of developmental landscapes by optimal-transport analysis of single-cell gene expression sheds light on cellular reprogramming. biorXiv, October 2017.
\\
Regev et al. (2017). The Human Cell Atlas. biorXiv, May 2017.

\end{document}